\documentclass{IEEEcsmag}

\usepackage[colorlinks,urlcolor=blue,linkcolor=blue,citecolor=blue]{hyperref}
\expandafter\def\expandafter\UrlBreaks\expandafter{\UrlBreaks\do\/\do\*\do\-\do\~\do\'\do\"\do\-}
\usepackage{upmath,color}

\usepackage[numbers]{natbib}

\jvol{XX}
\jnum{XX}
\paper{X}
\jmonth{August}
\jname{}
\jtitle{}
\pubyear{2026}

\begin{document}


\title{The Specification Paradox: Rethinking Requirements Engineering in the Age of AI}

\author{Tassio Sirqueira}
\affil{Department of Computer Science, Institute of Mathematics and Statistics, Rio de Janeiro State University, Rio de Janeiro, Brazil}

\author{Jessica Faciroli}
\affil{Faculty of Economics, Rio de Janeiro State University, Rio de Janeiro, Brazil}

\markboth{The Specification Paradox}{Requirements Engineering in the Age of AI}

\begin{abstract}
The growing adoption of Large Language Models (LLMs) in Software Engineering has reinforced the expectation that coding activities can be largely automated. However, this perception may represent yet another historical search for a solution capable of eliminating the inherent challenges of software development. This article discusses the transition from a code-centered paradigm to Specification-Driven Development. We argue that artificial intelligence reduces some of the effort associated with writing source code, but it does not eliminate the complexity of developing professional software systems. Instead, it shifts this complexity toward domain understanding, requirements elicitation, specification development, validation, maintenance, and software evolution. Building on this perspective, we discuss the renewed centrality of Requirements Engineering, considering its implications for productivity and software quality, as well as risks associated with automation bias, ambiguity propagation, Specification Overfitting, and the accumulation of Specification Debt. Finally, we propose the Specification Paradox: the more capable artificial intelligence systems become at automatically generating software, the greater the dependence on correct, complete, verifiable, and explainable human-produced specifications. We conclude that the future of Software Engineering will depend not only on machines’ ability to generate code, but also on humans’ ability to correctly specify, evaluate, and evolve what is intended to be built.
\end{abstract}

\maketitle

\chapteri{T}he history of Software Engineering has been marked by recurring enthusiasm surrounding technologies presented as ``silver bullets'' for the challenges of building complex software. Yet, as argued by \cite{brooks1987}, software complexity derives largely from the abstract, evolving, and interconnected nature of software systems and cannot be eliminated through isolated technological advances.

Generative artificial intelligence (GenAI), particularly Large Language Models (LLMs), renews this debate. By generating code from natural-language descriptions, these systems increasingly automate implementation and shift human effort toward defining intentions, constraints, expected behavior, and validating generated artifacts. Rather than eliminating software complexity, AI therefore redistributes it toward domain understanding, Requirements Engineering (RE), specification, validation, and evolution.

This shift favors Specification-Driven Development (SDD), in which specifications increasingly become direct inputs to AI systems while source code becomes a generated artifact. However, requirements are neither static nor always fully formalizable \cite{sommerville2016,berry2002}; they evolve through negotiation, interpretation, and changing organizational and domain contexts. AI does not eliminate these uncertainties and may instead operationalize them.

Requirements Engineering consequently becomes more, rather than less, important. As specifications become operational inputs to automated systems, their completeness, consistency, traceability, contextualization, and verifiability increasingly affect the quality of generated software.

This relationship leads to what we call the \textit{Specification Paradox}: the more capable AI systems become at generating code, tests, and documentation, the greater the dependence on high-quality human-produced specifications. Automation reduces implementation effort while increasing the importance of correctly defining what should be built.

\begin{figure*}[ht!]
\centering
\includegraphics[width=\textwidth]{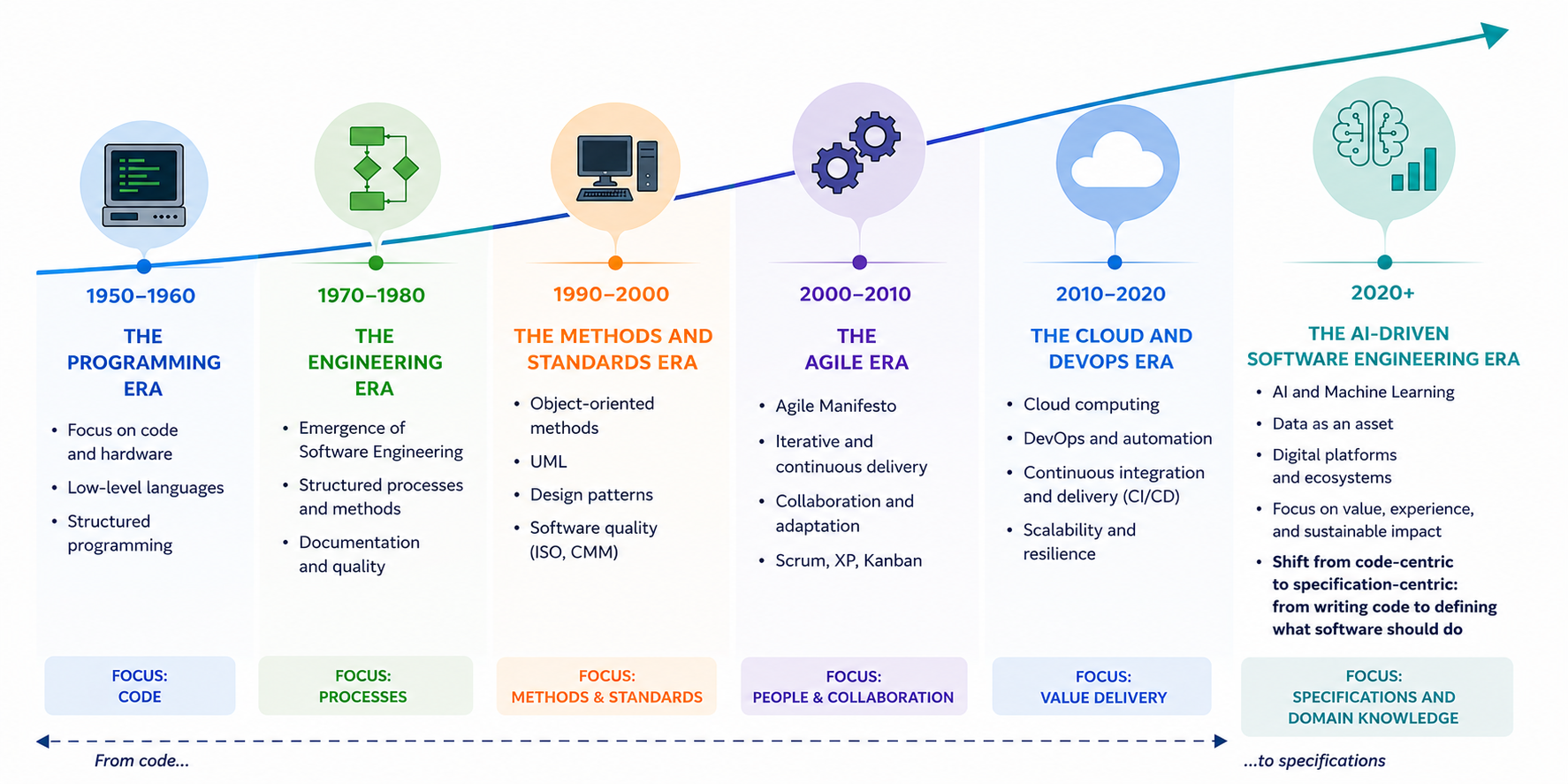}
\caption{Evolution of Software Engineering toward AI-driven development, in which engineering effort increasingly shifts from code implementation toward specifications and domain knowledge.}
\label{fig:evolution}
\end{figure*}

Figure \ref{fig:evolution} summarizes this transition from code-centered development toward specifications and domain knowledge as increasingly central engineering artifacts.

This article makes three contributions. First, it examines GenAI through Brooks' distinction between essential and accidental complexity \cite{brooks1987}. Second, it characterizes the transition from code-centered to specification-driven development and its implications for RE. Third, it introduces the \textit{Specification Paradox}, arguing that increasingly capable AI systems make high-quality human-produced specifications more consequential for software quality and alignment with domain needs.

\section{The Illusion of Automation (or ``The New Silver Bullet?'')}\label{ilusao}

\vspace{1em}

Successive advances in Software Engineering have raised the level of abstraction and reduced accidental complexity without eliminating the fundamental challenges of developing complex systems. GenAI follows this trajectory: LLM-based tools can rapidly generate code, tests, and documentation, reducing implementation effort while leaving activities such as problem understanding, requirements definition, architectural decision-making, and validation largely dependent on human expertise.

Moreover, apparent productivity gains may be partially offset by the effort required to understand, validate, integrate, maintain, and evolve AI-generated code \cite{11433212}. Automation therefore changes not only the amount of development effort but also where that effort is concentrated.

This shift can be understood through Brooks' distinction between accidental and essential complexity \cite{brooks1987}. AI can reduce accidental complexity by automating implementation tasks, but essential complexity remains rooted in the problem itself, including its domain constraints, requirements, and evolving objectives.

A further challenge is that fluent responses and plausible code may create an appearance of understanding, encouraging developers to overestimate model capabilities and overlook problems of consistency, verifiability, and alignment. Sophisticated outputs therefore still require critical human evaluation.

The emerging pattern is not the disappearance of software engineering work but its redistribution. As implementation becomes easier to automate, human effort shifts toward understanding the problem, specifying expected behavior, and validating generated artifacts. The central engineering challenge increasingly moves from how to write code toward how to correctly specify what should be built.

\section{From Code-Centered Development to Specification-Driven Development}\label{sdd}

\vspace{1em}

The trend illustrated in Figure \ref{fig:evolution} culminates in a fundamental shift in the focus of Software Engineering. Whereas source code has traditionally served as the primary engineering artifact, the growing capability of AI systems to generate implementations shifts the center of development toward specifications and explicit domain knowledge. As implementation becomes increasingly automated, human effort moves toward defining expected behavior, making constraints explicit, and validating generated artifacts.

This transformation favors the adoption of Specification-Driven Development (SDD) \cite{10.1007/978-3-540-24853-8_12}. Although specifications as central development artifacts are not a new idea, AI systems' ability to interpret natural-language descriptions expands their potential operational role. Specifications are no longer merely documents supporting communication among people; they can become inputs directly consumed by computational agents.

The underlying vision of SDD is straightforward: if a specification is sufficiently precise, complete, and verifiable, implementation can be largely automated. Source code then becomes a derived artifact, while specifications become central to defining and evolving software behavior. Changes and corrections can be made at the specification level and propagated to implementation through automated mechanisms.

This vision has parallels with Formal Methods and \textit{correctness by construction} approaches \cite{976937}, which have long sought to improve software reliability through higher levels of abstraction and rigorous specifications. A key difference is that contemporary AI systems can operate on less structured representations, including natural-language requirements, user stories, acceptance criteria, and technical documentation. This capability lowers some barriers to specification-driven practices but does not provide the semantic guarantees associated with formal specifications.

This distinction introduces a fundamental challenge. Natural-language specifications remain subject to ambiguity, inconsistency, and multiple interpretations. Human developers can use domain knowledge, experience, and interaction with \textit{stakeholders} to resolve such uncertainties; AI systems depend on the information and context made available to them. Consequently, a plausible implementation may faithfully reflect an incorrect or incomplete interpretation of the intended behavior. Implementation quality therefore becomes increasingly dependent on specification quality.

Traceability also becomes more critical in AI-driven development. Trust in automatically generated software requires mechanisms linking requirements to generated artifacts and providing evidence for the decisions underlying the resulting implementation \cite{11563783}. Such mechanisms become particularly important when code, tests, and documentation may all derive from the same specifications.

Software maintenance and evolution likewise become increasingly dependent on specification quality. In conventional environments, outdated documentation could coexist with source code functioning as the de facto representation of the system. In SDD, obsolete or inconsistent specifications can propagate directly to generated artifacts. Specification management therefore becomes a strategic capability for maintaining and evolving AI-assisted software systems.

SDD thus represents a conceptual change in what constitutes the primary software artifact: engineering effort moves from manually producing implementations toward explicitly defining, validating, and evolving intended system behavior. Its success will depend not only on increasingly capable AI models, but also on the ability of organizations to create and maintain specifications suitable for reliable software generation.

\section{Rethinking Requirements Engineering}\label{engreq}

\vspace{1em}

As source code becomes less of a bottleneck in software development, Requirements Engineering assumes a strategic role throughout the software lifecycle. In environments where AI systems can rapidly generate implementations, the primary engineering challenge increasingly shifts from \textit{how to implement} toward \textit{what exactly should be implemented}.

Requirements have traditionally served as communication mechanisms among stakeholders and development teams, reducing ambiguity sufficiently to guide implementation. Yet they are inherently incomplete, evolutionary, and subject to negotiation, reflecting the technical, organizational, and social dimensions of software development \cite{7111892}. These characteristics become particularly consequential when requirements are consumed directly by AI systems.

LLMs do not independently negotiate requirements, infer implicit organizational objectives, or resolve conflicting stakeholder intentions. Their outputs depend on the specifications and contextual information provided to them. Experienced developers may identify inconsistencies, question contradictory requirements, and seek clarification during implementation; AI systems may instead produce plausible implementations consistent with flawed or incomplete specifications. In this sense, AI does not eliminate ambiguity; it transforms ambiguity into executable behavior.

Consequently, requirements cease to be exclusively communication artifacts and increasingly assume a computational role. Specifications become an intermediate representation between stakeholder intentions and AI-generated software. This role increases the importance of properties such as consistency, verifiability, traceability, and explicit contextualization, because deficiencies in specifications can propagate directly across generated artifacts.

Traditional Requirements Engineering practices therefore acquire renewed importance. Structured elicitation, domain modeling, objective acceptance criteria, usage scenarios, requirement-derived tests, and traceability mechanisms become essential mechanisms for controlling the interaction between human engineers and AI systems. The growing ability to automate implementation consequently increases, rather than reduces, the need for rigorous RE.

The resulting reversal is central to the \textit{Specification Paradox}: as AI becomes more capable of generating code, tests, and documentation, the explicit knowledge of the problem and the specifications that represent it become increasingly important engineering assets. Requirements Engineering thus evolves from a predominantly supporting activity into a central mechanism for controlling the quality and alignment of AI-generated software.

\begin{figure*}[ht!]
\centering
\includegraphics[width=.92\textwidth]{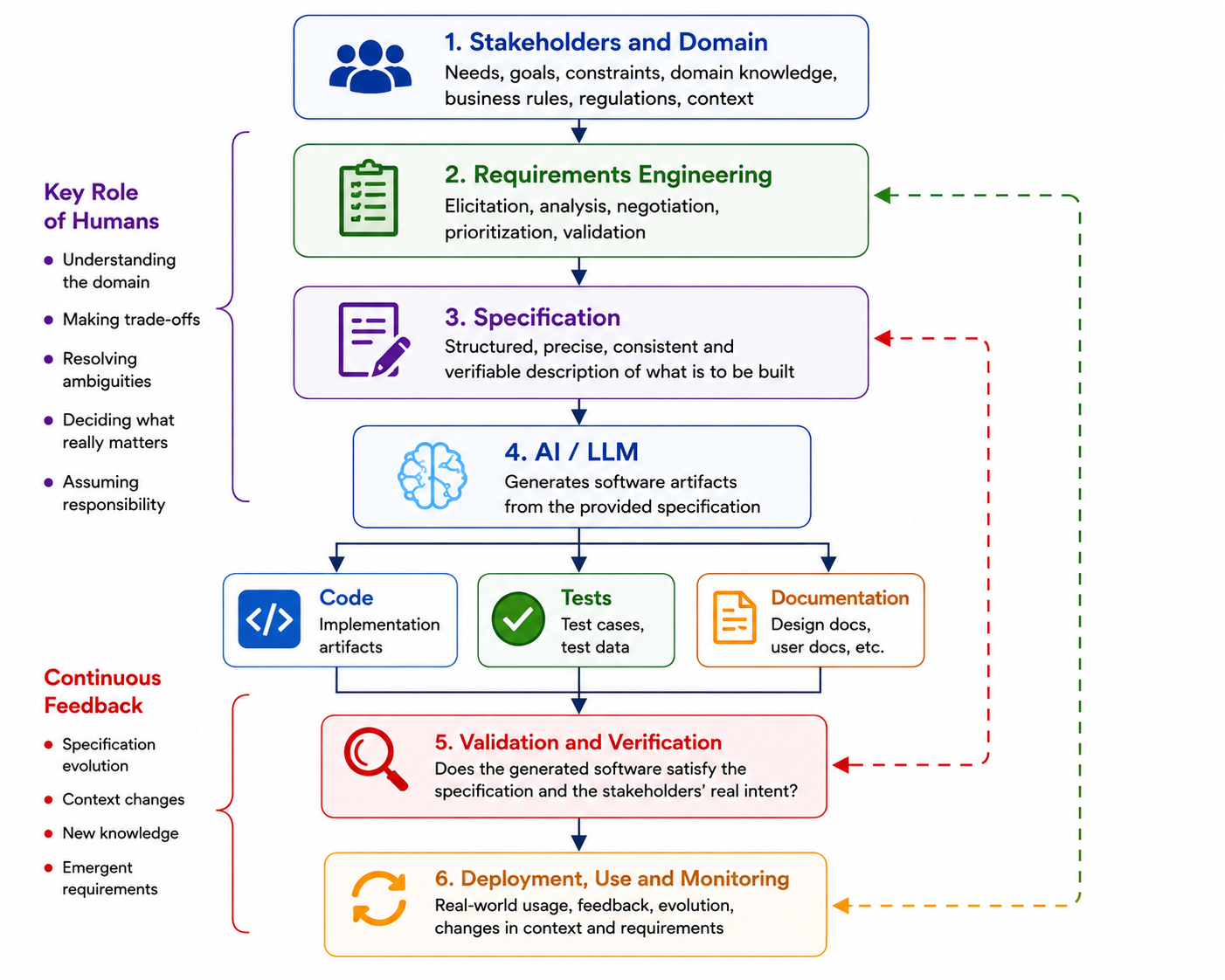}
\caption{
Conceptual model of SDD in the age of AI. Human stakeholders define requirements and domain knowledge that are refined into explicit specifications. AI systems generate software artifacts—including source code, tests, and documentation—from these specifications. Human engineers remain responsible for validation, verification, and continuous refinement as software evolves. The model illustrates that automation does not eliminate engineering effort; instead, it shifts it from implementation toward specification and validation.
}
\label{fig:sdd}
\end{figure*}

Figure \ref{fig:sdd} summarizes this transition. Specifications mediate stakeholder intentions and AI-generated artifacts, while human validation and feedback provide the mechanisms needed to preserve alignment as the software evolves.

\section{Recent Empirical Evidence}\label{hoje}

\vspace{1em}

Recent empirical evidence increasingly supports the conceptual shift illustrated in Figure \ref{fig:sdd}. Although AI systems can substantially accelerate implementation, the available studies suggest that productivity gains remain uneven and that generated artifacts still require extensive human validation.

A systematic review by \cite{10.1145/3809494}, covering 39 studies published between 2014 and 2024, found that most report productivity gains associated with reduced implementation time and the automation of repetitive tasks. However, approximately 59\% of the studies were exploratory, and only 15\% evaluated more than three dimensions of the SPACE productivity framework, indicating that evidence of sustained productivity improvements at project and team levels remains limited.

Productivity gains also do not necessarily imply improvements in software quality. AI-generated code may contain logical errors, architectural inconsistencies, and security vulnerabilities, while apparently correct implementations may fail under complex scenarios or implicit domain requirements. As emphasized by \cite{10.1145/3809494}, current evidence does not establish that AI-generated code is consistently more correct, secure, or maintainable. Faster implementation therefore increases the importance of validation and verification.

Specification quality and contextual information also strongly influence model performance. \cite{10628503} demonstrate this effect in RE by comparing isolated-sentence analyses with paragraph-level context. In their experiments, GPT-4 accuracy increased from approximately 41\% to 81\%, GPT-3.5 from 30\% to 63\%, and Mixtral from 33\% to 69\%. Although conducted in regulatory compliance, the study provides quantitative evidence that richer semantic context can substantially improve model interpretations.

The distribution of human effort also changes. Using eye tracking and interaction monitoring, \cite{10714560} found that developers working with LLM-generated code devote substantial cognitive effort to understanding, inspecting, validating, and correcting generated solutions. Automatic generation therefore does not eliminate human work; it shifts part of that work from implementation toward supervision and evaluation.

The adoption of LLMs also introduces new engineering challenges. An analysis of more than 29,000 developer questions on OpenAI forums identified recurring difficulties involving prompt engineering, API integration, architecture, deployment, debugging, and maintenance \cite{10.1145/3715007}. These findings indicate that AI does not remove software-development complexity but redistributes it across new technical activities.

Taken together, the available evidence suggests that AI can accelerate well-defined implementation tasks while increasing the importance of context, validation, and human oversight. These findings provide empirical support for the \textit{Specification Paradox}: as implementation becomes easier to automate, the reliability of AI-assisted software increasingly depends on the quality of the specifications and contextual information used to guide generation.

\subsection{Risks and Limitations}\label{riscoselimitacoes}

Recent advances in generative artificial intelligence do not eliminate longstanding Software Engineering challenges; instead, they introduce new ways in which specification deficiencies can affect software quality.

One such risk is what we refer to as \textit{Specification Overfitting}. Analogous to overfitting in Machine Learning, this phenomenon occurs when an AI-generated implementation closely satisfies the explicit specification but fails to address the broader problem, business needs, user expectations, or scenarios omitted during requirements elicitation. The resulting software may therefore be correct with respect to the specification while remaining incorrect with respect to the problem it is intended to solve.

This risk is amplified by the dependence of LLMs on the information provided to them. Omissions, ambiguities, and inconsistencies in specifications can be directly translated into software behavior rather than remaining merely documentation problems.

A related concern is \textit{automation bias}. Because AI-generated code is often fluent, readable, and apparently well structured, developers may overestimate its correctness and reduce critical scrutiny. Such plausibility can make semantic defects particularly difficult to identify, reinforcing the need for systematic human validation.

Automation may also amplify ambiguity propagation. In highly automated environments, a single interpretation of an ambiguous requirement can be rapidly materialized across code, tests, and documentation, propagating the same incorrect assumption through multiple software artifacts.

A further risk is what we refer to as \textit{Specification Debt}. Inspired by the concept of \textit{Technical Debt} \cite{10109339}, Specification Debt represents the accumulation of incomplete, inconsistent, redundant, poorly traceable, or insufficiently contextualized specifications whose deficiencies increase the cost and difficulty of generating, maintaining, and evolving software.

Unlike traditional technical debt, which typically becomes visible during implementation or maintenance, Specification Debt can affect the development process before code generation begins. When specifications become primary inputs to AI systems, their deficiencies may be systematically reproduced across generated artifacts and successive software versions. Low-quality specifications therefore cease to affect only communication among stakeholders and directly influence the behavior and evolution of generated software.

This shift also increases the importance of traceability and explainability. It is no longer sufficient to link requirements to source code and test cases; engineers may also need to determine which specification elements and contextual information influenced particular generated decisions and what evidence supports the resulting behavior.

Finally, automation creates an organizational risk: the ease of generating code may encourage reduced investment in requirements elicitation, modeling, and validation. Yet greater automation increases, rather than reduces, the importance of mature Requirements Engineering and quality-assurance practices. The central challenge is therefore not simply whether AI can generate software, but whether organizations can produce and maintain specifications robust enough to guide that generation reliably.

\subsection{Practical Implications for Software Engineers}\label{praticas}

The increasing incorporation of artificial intelligence into software development redefines, rather than reduces, the role of skilled software engineers. As implementation becomes increasingly automated, professional expertise shifts toward problem understanding, domain modeling, requirements definition, architectural reasoning, and validation of AI-generated solutions. Specifications consequently become living engineering artifacts that are continuously refined throughout the software lifecycle.

A first implication is the need to invest in specification quality. Requirements should be clear, consistent, verifiable, and sufficiently contextualized. Acceptance criteria, business constraints, alternative scenarios, and exceptional cases become essential inputs for guiding AI systems and evaluating whether generated solutions reflect the intended behavior.

Traceability also becomes increasingly important. AI-assisted development requires explicit relationships among requirements, architectural decisions, generated code, tests, and documentation. Such links support the validation of generated artifacts and help identify the specification elements or decisions associated with unexpected software behavior.

Testing and technical reviews must likewise adapt. Requirement-derived tests can verify whether AI-generated behavior corresponds to stakeholder intentions, while reviews should increasingly examine specifications, acceptance criteria, and design decisions before generation. Detecting a deficient requirement before it propagates across multiple generated artifacts is preferable to correcting those artifacts individually.

These changes also have implications for software engineering education. Programming and implementation skills remain fundamental, but RE, domain modeling, stakeholder communication, software architecture, systems thinking, and critical evaluation of AI-generated artifacts become increasingly important competencies.

Developing effective specifications for language models should also not be confused with simply writing \textit{prompts}. A prompt typically guides a particular interaction with a model, whereas a software specification must remain consistent, verifiable, traceable, and evolvable throughout the software lifecycle. Treating specification as prompt engineering risks producing solutions that are difficult to maintain and poorly aligned with established Software Engineering practices.

Finally, AI adoption should complement rather than replace mature engineering processes. Organizations with strong practices in RE, architecture, testing, and quality assurance are better positioned to benefit from automation. AI can amplify the consequences of both strong and weak engineering practices, making the quality of the artifacts that guide generation increasingly consequential.

\section{Conclusion}\label{conclusao}

\vspace{1em}

Generative artificial intelligence is transforming Software Engineering by automating the generation of code and other software artifacts. Yet it does not eliminate the essential complexity described by \cite{brooks1987}; rather, it redistributes engineering effort from implementation toward domain understanding, specification, validation, and software evolution.

This shift underlies the \textit{Specification Paradox}: the more capable AI systems become at generating software, the greater the dependence on high-quality human-produced specifications. As specifications increasingly guide automated generation, their quality directly affects the reliability and alignment of generated artifacts. Rather than diminishing RE, AI may therefore strengthen its strategic role.

This paradox has implications for research and practice, particularly for specification quality, verification, traceability, explainability, and evolution. Effective AI adoption consequently requires not only increasingly capable models but also mature engineering processes and professionals capable of defining and validating what those models should build.

Software Engineering thus continues to face a longstanding challenge: understanding the problem remains essential to constructing the right solution. If previous generations searched for a ``silver bullet,'' the AI era risks searching for a ``philosopher's stone'' capable of transforming imperfect specifications into perfect software. The more promising path lies in combining AI-driven automation with human responsibility for defining, validating, and evolving software behavior.

Ultimately, the future of software development may be defined not by the ability of machines to write code, but by our ability to correctly specify, validate, and evolve what we intend them to build.

\vspace*{-8pt}

 \section{ACKNOWLEDGMENTS}
This work was supported by the Brazilian funding agencies Coordenação de Aperfeiçoamento de Pessoal de Nível Superior (CAPES - Brazil) and 
Fundação Carlos Chagas Filho de Amparo à Pesquisa do Estado do Rio de Janeiro (FAPERJ - Brazil). The Article Processing Charge (APC) for the publication of this research was funded by the Coordenação de Aperfeiçoamento de Pessoal de Nível Superior - Brasil (CAPES) (ROR identifier: 00x0ma614). The authors used ChatGPT to assist with linguistic and grammatical review, as well as image enhancement. The authors are responsible for the entire content of the article.

\bibliographystyle{plainnat}
\bibliography{references}

\begin{IEEEbiography}
{Tassio Sirqueira} is a Professor at the Rio de Janeiro State University (UERJ), Department of Informatics and Computer Science, Institute of Mathematics and Statistics, Brazil. He received the Ph.D. degree in Computer Science from the Pontifical Catholic University of Rio de Janeiro (PUC-Rio). His research interests include Software Engineering, software quality, software evolution and aging, artificial intelligence, multi-agent systems, and data provenance. He has coordinated and participated in research projects involving software engineering, data analytics, and intelligent systems. He is also a reviewer and editor for scientific journals and conferences in Software Engineering and is a member of the Brazilian Computer Society (SBC).
\end{IEEEbiography}

\begin{IEEEbiography}
{Jessica Faciroli} is a professor at the Rio de Janeiro State University (UERJ) and a permanent faculty member of the Graduate Program in Economics (PPGCE-UERJ). She holds a Bachelor's degree in Economics from the Federal University of Juiz de Fora (UFJF) (2015), a Master's degree in Applied Economics (2018), and a Ph.D. in Applied Economics (2023), both from UFJF. She has experience in evaluating the impacts of social policies, labor market participation with an emphasis on gender and race, social network analysis, statistical and econometric methods, and the management and analysis of large administrative datasets. She is currently a Young Scientist of Our State (JCNE/FAPERJ) fellow for the 2026–2028 period.
\end{IEEEbiography}

\end{document}